\documentclass[11pt]{article}

\usepackage{graphicx}
\usepackage{dcolumn}
\usepackage{bm}
\usepackage{psfrag}

\title{A simple microscopic model for the dynamics of adhesive failure}
\author{Dominic Vella and L. Mahadevan
\\ \small Division of Engineering and Applied Sciences, Harvard University, \\ \small 29 Oxford Street, Cambridge MA 02138, USA}

\begin{document}
\maketitle
\begin{abstract}
 We consider a microscopic model for the failure of soft adhesives in tension based on ideas of bond rupture under dynamic loading. Focusing on adhesive failure under loading at constant velocity, we demonstrate that bi-modal curves of stress against strain may occur due to effects of finite polymer chain or bond length and characterise the loading conditions under which such bi-modal behaviour is observed. The results of this analysis are in qualitative agreement with experiments performed on unconfined adhesives in which failure does not occur by cavitation.
\end{abstract}

\section{\label{introduction} Introduction}

\begin{figure}
\centering
\includegraphics[height=6cm]{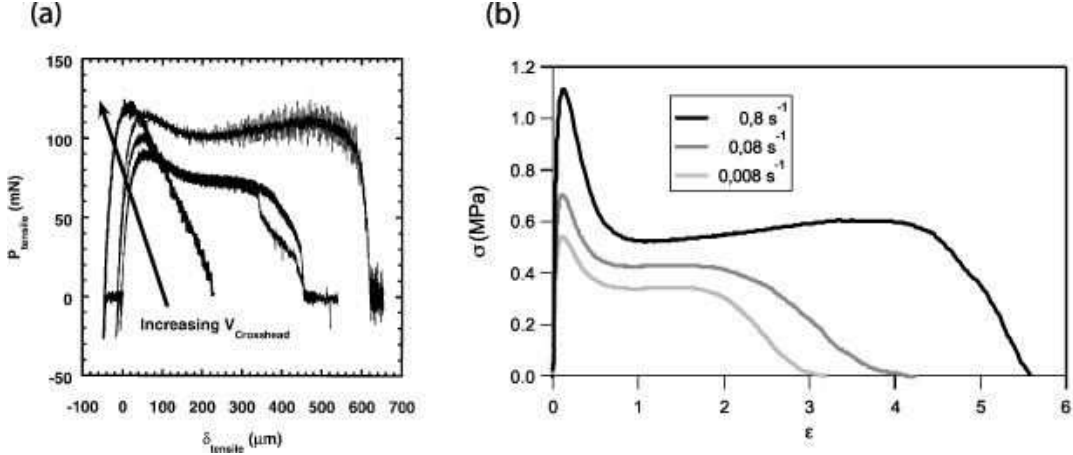}
\caption{Some typical tack curves showing both bi-modal and uni-modal
  behaviours. In both figures, the change from uni-modal to bi-modal
  is induced by changing the velocity of loading. (a) Loading force,
  $P_{tensile}$, as a function of imposed displacement,
  $\delta_{tensile}$, in the case of an unconfined adhesive (aspect
  ratio of order 1), taken from Crosby and Shull
  \cite{Crosby1999}. (b) Nominal stress, $\sigma$, as a function of strain, $\epsilon$, for a confined adhesive (a very thin layer of adhesive), taken from Chiche \cite{Chiche2003}.}
\label{crosby}
\end{figure}

The failure of natural and artificial adhesives is a subject of great importance involving the confluence of physical chemistry as well as statistical and continuum mechanics, but is as yet poorly understood. This is because of the multiple hierarchy of spatial and temporal scales involved, from the failure of individual molecular  bonds to that of the macroscopic sample, that occur over temporal scales that can also span many orders of magnitude. From a practical viewpoint, a particularly simple, yet useful, method of probing the failure of an adhesive is to stretch it at constant speed and plot the nominal stress against strain (referred to as \textit{tack curves})  to deduce some insight into the physical processes involved in failure. In fig.\ \ref{crosby} we show some typical tack curves observed during the loading of an adhesive in an unconfined geometry  \cite{Crosby1999} and in a confined geometry   \cite{Chiche2003}. We see that typical tack curves are approximately linear at small strains (since adhesives respond approximately linearly to small deformations), but at larger strains the nominal stress peaks (as the adhesive becomes damaged and is less able to resist deformation) and falls to zero as the adhesive fails. As the parameters associated with the tack test, such as the rate of loading,  the degree of confinement (i.e.\ the aspect ratio of the adhesive) or the type of adhesive are varied, a qualitative change in the tack curves shown in fig \ref{crosby} may be observed in the appearance   of a plateau or even a second peak. Since the area under a tack curve is a measure of the toughness of the adhesive, understanding the processes involved in determining the qualitative transitions in the stress-strain curve such as the maximum stress, and the presence or absence of a second peak or plateau are clearly efficacious in designing systems with a prescribed toughness.  

Some previous work has focused on explaining the appearance of a plateau in the tack curves (fig.\ \ref{crosby}(b)) as being due to the appearance of cavitation bubbles within the bulk of a confined adhesive layer (see, for example, \cite{Gay1999}). However this picture does not account for the subsequent rise in force that is observed even in the absence of cavitation. 
An alternative view focuses on the implications of non-linear elasticity on the shape of tack curves. For example, Sizaire and Legat \cite{Sizaire1997} used numerical computations based on the FENE-CR constitutive relation to model the response of a viscous suspension of finite length polymer chains. More recently, Roos and Creton \cite{Roos2004} used experimental data to fit a non-linear constitutive equation based on the `slip-tube' model of polymer elasticity to data obtained via tack tests. While some of the qualitative features of the tack curves shown in fig.\ \ref{crosby} are reproduced by these models, neither accounts for the failure of the material via a reasonable description in terms of the rupture of bonds or equivalent molecular picture.

Here we attempt to remedy this omission by considering these molecular effects within a simple microscopic model of the failure process. In particular, we show how changes in the nature of the adhesive and the rate of loading may lead to bi-modal tack curves whose physical origin is the finite length of either adhesive bonds or the polymers within the adhesive. We also characterize the loading conditions under which these bi-modal curves might arise and show that these are consistent with experimental conditions.  Although our simple model neglects many effects that are undoubtedly of importance,  it sheds some light on the underlying physics and suggests experiments that might further elucidate both the understanding and tailoring of adhesive toughness.

\section{\label{springs} A Microscopic Model}

The idea of bond rupture enhanced by the presence of an applied load was first used in the polymer literature  to explain the frictional properties of elastomers by  Schallamach \cite{Schallamach} following the classical work of Eyring \cite{Eyring} and Kramers \cite{Kramers}  on reaction rate theory. Similar ideas have also been used in more recent work on adhesion \cite{Chaudhury1999,Ghatak2000} to investigate the rate dependence of adhesive properties such as the work of adhesion, and also   to understand the molecular and cellular aspects of biological adhesion \cite{Bell1978}, \cite{Seifert2000}. A reduced model based on such ideas does not attempt to include the intricacies of the experimental geometry but allows us to focus on the physical processes that might lead to bi-modal tack curves.

\begin{figure}
\centering
\includegraphics[height=5cm]{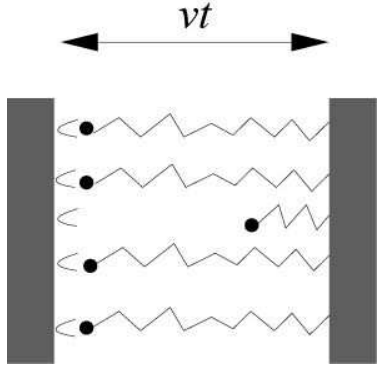}
\caption{Schematic of the model used in this discussion (adapted from Seifert \cite{Seifert2000}). Two rigid plates are pulled apart at constant velocity $v$ so that the loading of the springs connecting them is under displacement control. The bonds dissociate at a rate dependent on the load applied to them but are not allowed to reform.}
\label{setup}
\end{figure}

Our starting point is to consider two rigid plates connected by $N_b(t)$ non-linear springs of natural length $L$ (see fig.\ \ref{setup}). Chaudhury \cite{Chaudhury1999} and Ghatak \textit{et al.} \cite{Ghatak2000} have considered a similar model in which the springs represent the adhesive bonds that join the polymer adhesive to the plates. This model of adhesive failure (failure at the interface between the adhesive and apparatus) could, however, be applied more generally to cohesive failure (failure within the bulk of the adhesive) when the cohesive failure occurs via the rupture of polymer chains within the bulk rather than the appearance of cavities. We shall not, however, distinguish between these two modes of failure, since both can be represented by the schematic picture of fig.\ \ref{setup}.

Once the plates begin to be separated, the springs extend and so are stressed, causing them to dissociate at a rate that is dependent on the load to which they are subjected. The  dissociation rate is enhanced due to the effective lowering of the bond-potential barrier brought about by the application of a force. Assuming a simple Arrhenius type dependence, which is often referred to as the Bell model \cite{Bell1978}, we write the dissociation rate as:
\begin{equation}
\label{seifertrate}
k_-(t)=k_-^0\exp\left(\frac{F(t)\gamma}{k_BT}\right)
\end{equation} where $\gamma$ is a lengthscale associated with the bond potential, $k_B T$ sets the energy scale, $F(t)$ is the force with which each bond is loaded (assuming that they are all equally stressed) and $k_-^0$ is the dissociation rate in the absence of a force, which itself can be estimated in terms of the Eyring \cite{Eyring} or Kramers \cite{Kramers} theory of chemical reactions.
%
Given a particular constitutive relation for the non-linear springs it is a simple matter to write the corresponding evolution equation for the number of intact springs, $N_b$, remaining at time $t$. However, before doing so, we need a model for the response of the individual springs or polymer chains at the interface or in the bulk of the adhesive.   We choose a form that is simple for the purposes of analysis, but incorporates the two features that we desire here: linear behaviour at small displacements and a simple divergence in the force required to stretch the polymer as its length approaches the natural length, $L$. One possibility is the choice of a constitutive relation similar to the freely jointed chain model for stretching of a polymer chain given by  the inverse Langevin function (see, for example, \cite{Rubinstein}),  although this has no closed form. Therefore, for simplicity, we assume that the force required to stretch each spring by a distance $x$ is $F= k_BT x/b(L-x)$, where $L$ is some natural length, such as the contour length of the polymer chain between crosslinks, $b$ is a molecular length associated with the bond/polymer chain  such as the monomer or blob size  \cite{Rubinstein}  and  $k_BT$ is the thermal energy scale. The typical stiffness here has been chosen to reproduce that of a Gaussian chain \cite{Rubinstein} at small strains. In fact, the above force-displacement relation  is similar to a simple Pad\'{e}-like interpolation of the linear behavior for small displacements and the divergence implicit in the inverse Langevin function \cite{Pade}.

Assuming that the loading of bonds/polymer chains takes place at the same constant velocity as the plates are separated, $v$, the constitutive law described above allows the force on each remaining bond to be written as a function of time, $t$:
\begin{equation}
F(t)=\frac{k_BT}{bL}\frac{vt}{1-\frac{vt}{L}}
\label{dynforce}
\end{equation} 
Here we have assumed that the rate of reforming of bonds is relatively small, which is certainly true if the pulling force is moderate or large, the case of most interest here. There are three natural timescales in this problem. The first of the time-scales is the mean lifetime of an unforced spring, namely $1/k_-^0$. The second characterises the rate at which energy is stored in a spring compared to thermal energy $bL/\gamma v$ and provides us with a useful non-dimensionalisation of time: $\tau\equiv \gamma vt/ bL$. The third is simply the time taken to stretch a constituent spring to its natural length $L$, given by $L/v$. From the
ratios of these time-scales, we find two dimensionless parameters that govern the evolution of the system. The first of these is $\epsilon \equiv \gamma v/k_-^0bL$, the ratio of the intrinsic off rate to the loading rate, which is readily varied by varying the rate at which loading takes place, $v$. The second dimensionless parameter
is the time at which the natural length is reached, $\tau_m \equiv \gamma/b$, and is the ratio of the width of the bond-potential barrier and the monomer length. As this is dependent solely on the chemical properties of the adhesive, it is more difficult to vary experimentally, though this may, in principle, be achieved by altering the typical monomer bond length, $b$, and the strength of the monomer-monomer interaction; large values of $\tau_m$ correspond to very soft (long range) interactions between small monomers, such as Coulomb-type interactions between nano-particles, while small values of $\tau_m$ correspond to short range interactions between large monomers, such as van der Waals interactions between macromolecular globules. Making use of (\ref{seifertrate}) and (\ref{dynforce}) the governing equation for the number of intact bonds, i.e. $dN_b/dt = -k_-(t) N_b$, may be written in dimensionless form as
\begin{equation}
\frac{dN_b}{d\tau}=-\frac{1}{\epsilon}\exp\left(\frac{\tau}{1-\tau/\tau_m}\right)N_b.
\label{tevolve}
\end{equation} The change of variable $u=\tau/(1-\tau/\tau_m)$ allows us to write the solution of (\ref{tevolve}) in the form
\begin{equation}
N_b(\tau)=N_b(0)\exp\left(-\frac{1}{\epsilon}\int_0^{u(\tau)}\frac{\exp(w)}{(1+ w/\tau_m)^2}dw\right),
\end{equation} which can then be expressed in terms of the exponential integral $Ei(z)$, although this representation of the solution is of little interest in what follows. Here   $N_b(0)=N_0$, the initial number of intact bonds; however, we take $N_0=1$ so that $N_b(\tau)$ represents the fraction of the original bonds remaining at time $\tau$.  Then we may write  the total dimensionless force on the plates $\cal{F}(\tau)$ required to maintain the constant velocity displacement (scaled by $\gamma/k_BT$)   as the product of the force per bond and the number of intact bonds so that
\begin{equation}
{\cal F}(\tau)= \frac{\tau}{1-\tau/\tau_m}N_b(\tau).
\label{pullforce}
\end{equation}

\begin{figure}
\centering
\includegraphics[height=5cm]{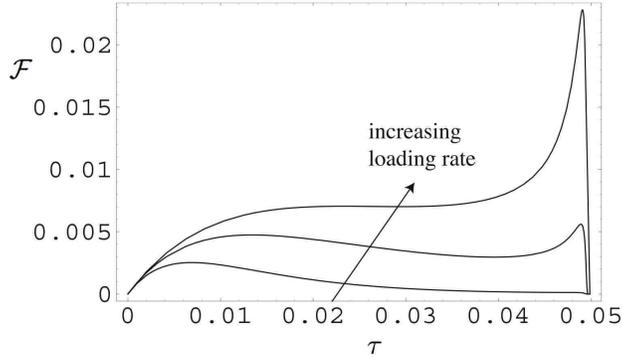}
\caption{Typical dimensionless force-displacement graphs for the model shown in Figure 2. As the rate of  loading is increased (in the sense indicated by the arrow), the tack curves change from unimodal to bimodal and then back to unimodal. Here $\tau_m=0.05$ and the rate  of loading is proportional to the parameter $\epsilon$, which takes  values $0.006$ (unimodal with a slow force decay preceding failure), $0.01$ (bimodal) and $0.013$ (unimodal with a very sharp force drop preceding failure).}
\label{forcedisp}
\end{figure}

As a function of time (or equivalently of displacement) the force   (\ref{pullforce}) is not monotonic  but instead rises slowly  and then decays down to zero. An explanation for an intermediate maximum follows from the simple fact that initially the force per bond increases with the displacement so that the total force also increases. However, as individual bonds start to break in response to the applied load, the force eventually starts to decrease. What is more interesting is that for certain values of the parameters $\epsilon$ and $\tau_m$, it is possible for there to be two maxima in this curve. Examples in which there are one or two maxima in the force are given in fig.\ \ref{forcedisp}.

Having observed that it is possible to obtain bi-modal tack curves that bear some qualitative resemblance to those shown in fig.\ \ref{crosby} (a), we now focus on understanding such curves in more detail. 

\section{Phase space for bi-modal tack curves}

Next, we quantify the regions of $(\tau_m,\epsilon)$ space in which bi-modal tack curves occur and then explain  the physical origin of this behaviour. We first obtain an expression for the (non-dimensional) times, $\tau^*$, at which ${\cal F}(\tau)$ has a turning point. Differentiating the expression for $\cal{F}(\tau)$ in (\ref{pullforce}) and using (\ref{tevolve}) to eliminate $dN_b/d\tau$ we see that the $\tau^*$ satisfy
\begin{equation}
\epsilon=\tau^*(1-\tau^*/\tau_m)\exp\left(\frac{\tau^*}{1-\tau^*/\tau_m}\right)\equiv G(\tau_m,\tau^*).
\label{genteqn}
\end{equation} Although this equation does not admit an analytic solution, the general properties of the function $G(\tau_m,\tau)$ are enough for our purposes. Clearly $G(\tau_m,0)=0$ and as $\tau \rightarrow \tau_m$, the value of $G(\tau_m,\tau)$ diverges exponentially. However, in the interval $(0,\tau_m)$, $G$ may have, depending on the value of $\tau_m$, either zero or two turning points. Assuming that there are two turning points, they occur when $\tau=\tau_{\pm}$ where
\begin{equation}
\tau_\pm=\frac{3\tau_m-\tau_m^2\pm \tau_m\sqrt{\tau_m^2-6\tau_m+1}}{4}.
\label{turnpoints}
\end{equation} For the values of $\tau_\pm$ to be real and distinct we require that $\tau_m^2-6\tau_m+1>0$, i.e. $\tau_m<3-2\sqrt{2}$ or $\tau_m>3+2\sqrt{2}$. For values of $\tau_m$ satisfying the latter condition, it is easy to see that $\tau_+<0$ which is not physically meaningful so that we consider only the first possibility. In this region $0<\tau_\pm<\tau_m$ so that we really do see both of the turning points in the function $G(\tau_m,\tau)$ in the interval $[0,\tau_m)$. From the geometrical constraints on $G$, we see immediately that $\tau_-$ corresponds to a local maximum and $\tau_+$ a local minimum in $G(\tau_m,\tau)$. Thus we see that  the applied force, ${\cal F}$, is bimodal only if $\epsilon \in (\epsilon_-,\epsilon_+)$ where $\epsilon _\pm=G(\tau_m,\tau_\mp)$.

\begin{figure}
\centering
\includegraphics[height=5cm]{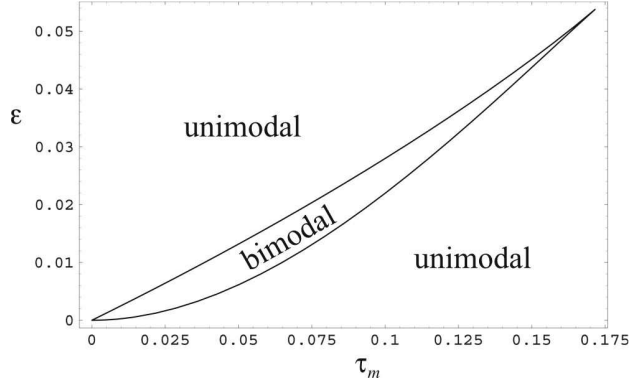}
\caption{Phase diagram showing the values of the parameters $\tau_m$, the scaled bond-potential width and $\epsilon$, the scaled loading rate for which the scaled applied force, $\cal{F}(\tau)$, is unimodal or bimodal. As expected, for very large or very small loading rates, the tack curve is unimodal, but for a narrow range of loading rates, bimodal tack curves can exist.}
\label{phasediag}
\end{figure}

This result is most easily summarized by means of a phase diagram in $(\tau_m,\epsilon)$ phase space showing where we expect the two types of force curves (unimodal and bimodal) to occur. The borders of these regions can be calculated analytically from the preceding analysis and are shown in fig.\ \ref{phasediag}, which shows that there is only a very small range of the parameters $\tau_m$ and $\epsilon$ for which the bimodal behaviour can be observed.

Comparison with experiments requires estimates for the typical values of the parameters $\epsilon$ and $\tau_m$. Assuming that the failure occurs as a result of the rupture of individual polymer chains within the bulk rather than adhesive bond rupture at the interface we use the result for a Gaussian chain \cite{Rubinstein} which has the constitutive response given by (\ref{dynforce}) with $L=Nb$ where $N\sim 10^8$ is the number of monomers in an individual chain and $b\sim10^{-9}$m is the typical monomer  (or Kuhn) length.  Along with $\gamma\sim10^{-10}$m, this gives $\tau_m \sim 0.1$. It is difficult to estimate a typical value of $\epsilon$ because of the ambiguity in $k_-^0$. However, it is enough to note that whatever the value of $k_-^0$, the typical value of $\tau_m$ is consistent with our model's prediction that the appearance of bimodal tack curves can be controlled by varying the velocity of pulling, as is observed experimentally.

To understand the behaviour exhibited in fig.\ \ref{phasediag} we first consider the existence of an upper bound on the value of $\tau_m$ ($=\gamma/b$) for which bimodal force displacement curves may occur. For larger values of $\tau_m$, the nonlinearity in the force law becomes important only once most of the bonds have already been broken and so no second peak is observed. For a given value of $\tau_m$ (sufficiently small to avoid the previous pitfall), very fast loading (large $\epsilon=\gamma v/k_-^0bL$) will cause most of the bond breakage to occur only as the strain-stiffening starts to become important and, again, only one force peak is observed. We thus expect that bi-modal tack curves can only occur below some critical loading rate $\epsilon_+$.  Notice that the bi-modal curves presented in the experimental data fig.\ \ref{crosby} are not observed for large loading rate: just as predicted by our analysis and simple physical mechanism. On the other hand, if we load the springs too slowly then very few of them will be broken by being stressed and instead break over their natural time-scale $1/k_-^0$ so that few remain by the time they are extended to the point at which strain hardening is appreciable. This leads us to expect that for a given value of $\tau$ there should be a minimum loading rate $\epsilon_-$ below which the force curve is again unimodal.

As we have just seen, bimodal curves occur when the individual bond-springs are extended towards their natural length (when $\tau = \tau_m$) and stiffen as a result. We observe that the decay of the applied force after this peak is much more rapid than the decay after the first peak. Informal comparison with the experimental results of Crosby and Shull \cite{Crosby1999}  presented in fig.\ \ref{crosby} (a) confirms that this is also observed experimentally. However, the results of Chiche \cite{Chiche2003} show that in a confined geometry where cavitation occurs, such a rapid decay is not experienced after the second peak. A natural question that arises is whether the rapid decay is a result of a particular choice of parameters or is an intrinsic characteristic of such a model. In the next section, we show that our model predicts that this decay in the force is \textit{necessarily} very rapid.

\section{Late time asymptotics}

To study the behaviour of the applied force after the second peak, we study the behaviour of ${\cal F}(\tau)$ as $\tau\rightarrow \tau_m$, i.e.\ for $\tau$ such that $ \tau_m-\tau  \ll 1$. In terms of the parameter $u \equiv \tau_m/(1-\tau/\tau_m)$ this is equivalent to the limit $u \gg 1$. With this substitution, we can integrate (\ref{tevolve}) so that, up to a constant of integration, the number of bonds $N_b$ satisfies
\begin{equation}
\log N_b =-\frac{\tau_m^2\exp(-\tau_m)}{\epsilon} \int_{\tau_m}^{u} \frac{e^w}{w^2} dw,
\end{equation} which can then be integrated by parts to give
\begin{equation}
\log N_b =-\frac{\tau_m^2\exp(-\tau_m)}{\epsilon}  \frac{e^u}{u^2} +{\cal O}\left(\frac{e^u}{u^3}\right).
\end{equation} Keeping only the leading order terms, recalling that $u \gg 1$, yields an expression for the force, which to leading order is
\begin{equation}
{\cal F}(\tau) \sim  \frac{\tau}{1-\tau/\tau_m} \exp\left( -\frac{\exp(-\tau_m)}{\epsilon}(1- \tau/\tau_m)^2\exp\left(\frac{\tau_m}{(1-\tau/\tau_m)}\right)\right).
\label{death}
\end{equation} It is immediately clear from this expression that when $\tau \approx \tau_m$, the pulling force decays super-exponentially and that the initial conditions are irrelevant since their influence is exponentially small. In fact, the decay given in (\ref{death}) is impossible to observe in the numerical solutions of (\ref{tevolve}) since it is important only \emph{very} close to the singularity, where the errors in the numerical integration scheme overwhelm the effect that we wish to observe.

This result shows that the behaviour of the force-displacement curve as $\tau\rightarrow \tau_m$ is qualitatively similar to that observed in some experiments (e.g. \cite{Crosby1999}), but not others (e.g. \cite{Chiche2003}),  which differ from each other mainly in the nature of the experimental geometry highlighting the role of confinement. In an unconfined adhesive, the failure is initiated by a fingering instability that occurs around the edge of the adhesive \cite{Crosby1999}; a confined adhesive will fail via cavitation within the bulk of the adhesive \cite{Chiche2003}. In the former case, the constitutive response of the elastomer that remains is unchanged but in the latter case, the properties of the adhesive change because of its ability to accommodate the imposed deformation via bubble growth. Thus our microscopic model     applies only to adhesive loading in the unconfined case. In this scenario, damage is accumulated by the adhesive as a whole even though the individual constituents remain largely undamaged. This notion of damage to the adhesive as a whole can be used to recast the microscopic model of \S 2  in terms of a continuum theory, to which we turn next.

\section{A Continuum Model}

To adapt the notion of adhesive failure in terms of the microscopic spring model of \S \ref{springs} to   the case of a continuous adhesive of constant volume we introduce a damage parameter, $1-\alpha$, which increases from $0$ to $1$ as damage is accumulated by the adhesive. Here $\alpha$ is defined as the ratio of the actual stress $\sigma$ required to impose a given displacement and the stress experienced by an \textbf{u}ndamaged adhesive subject to the same displacement, $\sigma_u$ i.e. $\sigma=\alpha \sigma_u$.

By analogy with the parallel spring model of \S \ref{springs}, we postulate an Arrhenius type evolution for $\alpha$ as a function of the imposed stretch, $\lambda$, namely
\begin{equation}
\frac{d \alpha}{d \lambda}=-\frac{\alpha}{\omega} \exp\left( \frac{\sigma_u V}{k_BT}\right),
\label{alphaeqn}
\end{equation}  whose   basis is  the correspondence between $\alpha$ and the number of unbroken springs  $N_b$ in the microscopic model. Here, $\omega\equiv\dot{\lambda}/k_-^0$ is the ratio of the stretch rate to the unforced dissociation frequency, the continuum analog of
$\epsilon$, and $V\sim\gamma l_e^2$ is some characteristic volume (where $\gamma$ is again a typical length associated with the bond potential and $l_e$ is the typical entanglement length or distance between crosslinks of the polymer network, so that stress is distributed over an area $l_e^2$). To complete the formulation of  the continuum model, we need a constitutive equation for the stress in terms of the stretch or equivalently a free energy $U_f(\lambda)$ such
that
\begin{equation}
\sigma_u=-\lambda \frac{\partial U_f}{\partial \lambda}.
\end{equation}
To account for  the finite length effects of the constituent chains in the adhesive, we use the form of the free energy, $U_f$, suggested by Gent (1996)
\begin{equation}
U_f=\frac{k_BT}{V'}J_m\log\left( 1-\frac{J}{J_m}\right), \label{Gent}
\end{equation} 
where $V'\sim l_e^3$ is the typical microscopic volume characteristic of the polymer network, $J(\lambda_x,\lambda_y,\lambda_z)\equiv\lambda_x^2+\lambda_y^2+\lambda_z^2-3$ (subject to the constraint $\lambda_x\lambda_y\lambda_z=1$ imposed by incompressibility) is the first invariant of the deformation gradient and $J_m=J(\lambda_m)$, for some $\lambda_m$, is the maximum value of this invariant corresponding to the maximal possible stretch of the adhesive, $\lambda_m$. We note that when $J\ll J_m$, the above free energy reduces to $U_f \simeq \frac{k_BT}{V'} J$,  the classical expression for a neo-Hookean solid. For an incompressible elastomeric adhesive loaded axially, as we will assume here,  $\lambda_x=\lambda_y$ so that $J\equiv J(\lambda)=\lambda^2+2\lambda^{-1}-3$. 

\begin{figure}
\centering
\includegraphics[height=5cm]{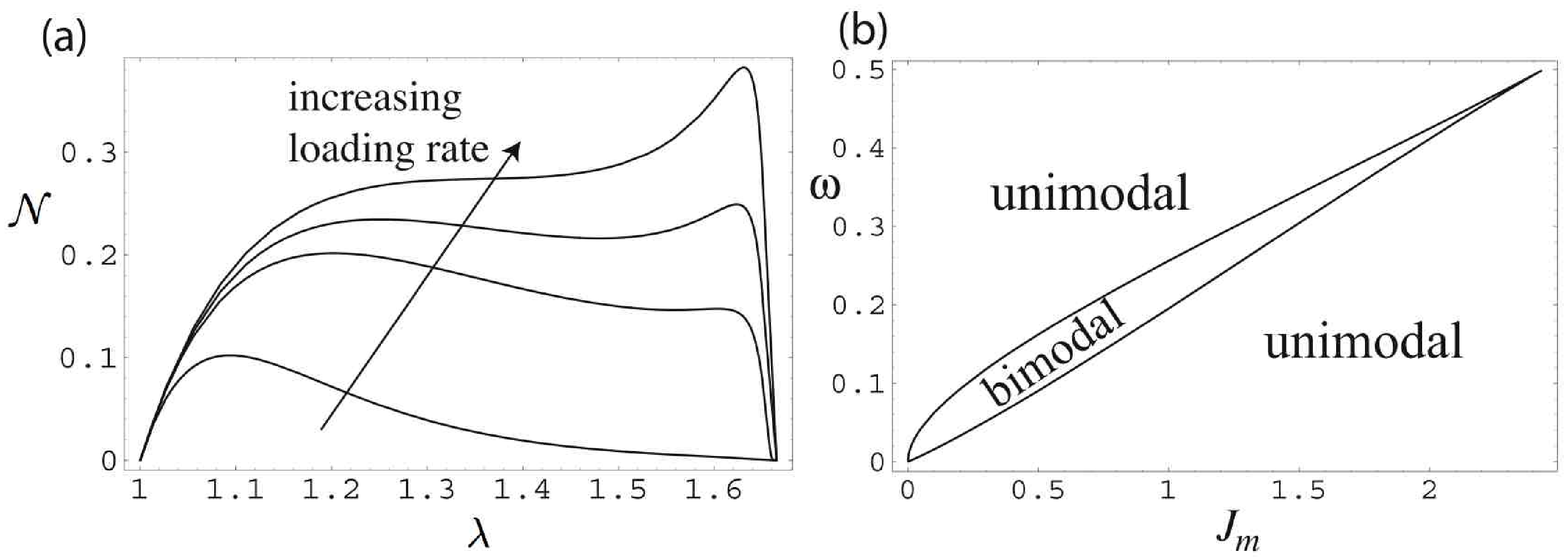}
\caption{The scaled nominal stress $N$ as a function of the applied stretch $\lambda$ obtained from the continuum model (\ref{alphaeqn}-\ref{Gent}). (a) Increasing the loading  rate, parametrized by $\omega$, changes the behaviour from unimodal  to bimodal and back again. Here $J_m=1$, $\nu=0.05$ and $\omega$  takes the values $0.1,0.2,0.23$ and $0.26$. (b) Numerically computed phase diagram showing the values of the scaled loading rate $\omega$ and the value of the first invariant at failure, $J_m$, for which bimodal and unimodal behaviours are observed. Here the scaled bond-potential barrier $\nu \sim \gamma/l_e =0.05$.}
\label{contfigs}
\end{figure}

In this formalism, it is a simple matter to solve (\ref{alphaeqn}-\ref{Gent}) subject to the initial condition that $\alpha(\lambda=1)=1$ via numerical methods. Such a solution can then be used to infer the nominal stress, ${\cal N}$, given by
\begin{equation}
\mathcal{N}=-\alpha \frac{\partial U_f}{\partial \lambda} \label{N}.
\end{equation} 
Experimentally, this is the quantity most commonly measured and thus allows us to make some simple qualitative comparisons. In fig.\ref{contfigs} (a) we show representative solutions of
(\ref{alphaeqn}-\ref{N}) for parameter values $J_m=1$ and $\nu\equiv
V/V'\sim\gamma/l_e=0.05$ as the loading rate is increased. We note
that the macroscopic analog of the bond potential barrier $\nu$ is
characterized as a fraction of the entanglement length rather than the
monomer length as in \S2. With this caveat, we see that our results
are qualitatively very similar to those  obtained using the
microscopic model in \S2; in particular, there is a range of loading
rates for which the nominal stress is bimodal, and for very large
loading rates  the nominal stress decays very rapidly after the second
peak. It is also possible to characterize the region of
$(J_m,\omega,\nu)$ space that leads to bi-modal tack curves, although
this must be done numerically here. In fig.\ \ref{contfigs} (b) we
characterize this region and see results that are, again, qualitatively similar to those in \S 2.

\section{\label{conclusion} Conclusions}

The existence of a second peak in tack curves obtained in adhesive testing and failure is a common feature of many recent experiments. When combined with the observed reduction of the nominal cross-sectional area of the adhesive  at the point where the second peak occurs this indicates that the modulus of the adhesive as a whole must have increased by this point. In this paper, we have investigated the possibility that this might be produced by the effects of finite bond/chain length and shown that such finite length effects, in the absence of cavitation, produce tack curves in qualitative agreement with those obtained experimentally with unconfined adhesives (e.g. Crosby and Shull \cite{Crosby1999}).  Our model allows us to also determine the region in phase space where this bi-modal behavior may be seen, and further explains the rapid failure of the adhesive following the second peak in the force. We predict that these bi-modal features should disappear at high pulling velocity, as is observed in such experiments.

One deficiency in the model considered here is its exclusion of the flow-induced viscoelastic nature of the adhesive itself; for simplicity we have considered purely elastic effects, with the rate dependence coming in via a simple kinetic mechanism. This might simply be remedied by incorporating the viscous and elastic response of the polymer chain with a bond joining this chain to the substrate. Another improvement to the model presented here would be to take into account the spatially varying nature of the problem. There are several levels of sophistication at which such an approach could be introduced. Perhaps the simplest modification would be to consider an axially symmetric problem but to account for the variations in chain length within the adhesive that occur because of the necking phenomenon that is a result of incompressibility of the adhesive and the pinning of the contact region to the plate. Ultimately, with the full spatial dependence accounted for, it would be possible to investigate the interplay between the onset of the fingering instability and finite chain length effects. Additionally an important aspect of any continuum theory of failure that we have not accounted for is the role of stress concentrations in the vicinity of the crack that demarcates the failure region. This effect is important in brittle materials, but its relative role in elastomers is debatable since the locally high stresses lead to crack blunting and even slow flow in soft materials. In the absence of experimental studies of the  relative importance of crack blunting, we defer this important theoretical question to the future. 

Our work has focused on one possible mechanism for the appearance of bi-modal tack curves. Other effects that might enhance this could arise from the realignment of polymer domains within the adhesive: at intermediate displacements, it is energetically cheaper to accommodate the imposed deformation by rotation of polymer domains than to stretch them. This semi-soft response (see, for example Warner and Terentjev \cite{Warner} for a thorough discussion of semi-softness) alters the constitutive response of the material and would allow for a relaxation of nominal stress via domain rotation. Once the maximal rotation has occurred, the domains must once again deform allowing them to be stretched as springs before they eventually break thence giving rise to a second peak. Indeed, this suggests that experiments that probe the structural rearrangements of the adhesive (e.g. using birefringence measurements) simultaneously during mechanical testing  might help unravel the relative contributions of domain rotation and simple stretching mechanisms. 

\paragraph*{Acknowledgements}
We are grateful for many helpful suggestions made by M.~K.~Chaudhury,
A.~J.~Crosby and M.~L.~Roper. DV was supported by the Choate Fellowship at Harvard.


\begin{thebibliography}{18}

\bibitem{Bell1978} G. I. Bell, \textit{Science} \textbf{200}, 618 (1978).

\bibitem{Chaudhury1999} M. K. Chaudhury, \textit{J. Phys. Chem. B} \textbf{103}, 6562-6566 (1999).

\bibitem{Chiche2003} A. Chiche, \textit{D\'{e}collement d'un adh\'{e}sif mou: fracture et cavitation}, Ph.D. Thesis, Universit\'{e} Paris VII (2003). 

\bibitem{Crosby1999} A. J. Crosby and K. R. Shull, \textit{J. Polym. Sci. B: Polym. Phys.} \textbf{37}, 3455-3472 (1999).


\bibitem{Eyring} H. Eyring, \textit{J. Chem. Phys.} \textbf{4} 283 (1936).

\bibitem{Gay1999} C. Gay and L. Leibler, \textit{Phys. Rev. Lett.} \textbf{82} 936-939 (1999).

\bibitem{Gent1996} A. N. Gent, \textit{Rubber Chem. Technol.} \textbf{69}, 59-61 (1996).

\bibitem{Ghatak2000} A. Ghatak, K. Vorvolakos, H. She, D. L. Malotky and M. K. Chaudhury, \textit{J. Phys. Chem. B} \textbf{104}, 4018-4030 (2000).


\bibitem{Kramers} H. Kramers, \textit{Physica} \textbf{7} 284 (1940).

\bibitem{Pade} We note this is not a true Pad\'{e} approximant, which  would be anti--symmetric under the transformation $x \rightarrow -x$ and thus have the form  $F= k_BT x/(L^2-x^2)$; however, for our purposes here the simplest form suffices. Changing the form of the divergence to accomodate other microscopic theories such as the worm-like-chain model is possible, but does not change any of the results qualitatively.

\bibitem{Rubinstein} M. Rubinstein and R. Colby, \textit{Polymer Physics}, Oxford (2003).

\bibitem{Roos2004} A. Roos and C. Creton, \textit{Macromol. Symp.} \textbf{214}, 147-156 (2004).

\bibitem{Schallamach} A. Schallamach, \textit{Proc. Phys. Soc. B} \textbf{66}, 386 (1953).

\bibitem{Seifert2000} U. Seifert,  \textit{Phys. Rev. Lett.} \textbf{84}, 2750-2753 (2000).

\bibitem{Sizaire1997} R. Sizaire and V. Legat, \textit{J. Non-Newtonian Fluid Mech.} \textbf{71}, 89-107 (1997).

\bibitem{Warner} M. Warner and E.M. Terentjev,  \textit{Liquid Crystal Elastomers}, Oxford University Press (2003).

\end{thebibliography}
\end{document}